\documentclass[12pt]{iopart}
\usepackage{graphicx}
\usepackage{subfigure}
\usepackage{iopams}
\usepackage{tabularx}
\usepackage{enumerate}
\usepackage[colorlinks,linkcolor=blue,citecolor=blue,urlcolor=blue]{hyperref}
\begin{document}

\title{Exact non-Hermitian mobility edges and robust flat bands in two-dimensional Lieb lattices with imaginary quasiperiodic potentials}

\author{Xiang-Ping Jiang}
\address{Zhejiang Lab, Hangzhou 311121, China}
\ead{2015iopjxp@gmail.com}

\author{Weilei Zeng}
\address{Zhejiang Lab, Hangzhou 311121, China}

\author{Yayun Hu}
\address{Zhejiang Lab, Hangzhou 311121, China}
\ead{yyhu@zhejianglab.edu.cn}

\author{Peng Liu}
\address{Zhejiang Lab, Hangzhou 311121, China}
\address{College of Information Science and Electronic Engineering, Zhejiang University, Hangzhou 310027, China}
\ead{liupeng@zju.edu.cn}

\vspace{10pt}
\begin{indented}
\item[]
\end{indented}

\begin{abstract}
The mobility edge (ME) is a critical energy delineates the boundary between extended and localized states within the energy spectrum, and it plays a crucial role in understanding the metal-insulator transition in disordered or quasiperiodic systems. While there have been extensive studies on MEs in one-dimensional non-Hermitian (NH) quasiperiodic lattices recently, the investigation of exact NH MEs in two-dimensional (2D) cases remains rare. In the present study, we introduce a 2D dissipative Lieb lattice (DLL) model with imaginary quasiperiodic potentials applied solely to the vertices of the Lieb lattice. By mapping this DLL model to the 2D NH  Aubry-Andr{\'e}-Harper (AAH) model, we analytically derive the exact ME and find it associated with the absolute eigenenergies. We find that the eigenvalues of extended states are purely imaginary when the quasiperiodic potential is strong enough. Additionally, we demonstrate that the introduction of imaginary quasiperiodic potentials does not disrupt the flat bands inherent in the system. Finally, we propose a theoretical framework for realizing our model using the Lindblad master equation. Our results pave the way for further investigation of exact NH MEs and flat bands in 2D dissipative quasiperiodic systems.
\end{abstract}

%
%
%
\maketitle
%
%

\section{Introduction}
Anderson localization~\cite{anderson1958absence}, a phenomenon characterized by the absence of particle diffusion in disordered systems, has been extensively studied in the field of condensed matter physics. In three-dimensional (3D) systems, the Anderson localization transition can manifest as a consequence of varying disorder strength or energy. This transition is typically manifested as mobility edges (MEs), critical energy separates the localized states from extended states~\cite{lee1985disordered,kramer1993localization,abrahams201050,evers2008anderson}. The degree of localization is intricately linked to the spatial dimensionality of the system. According to the well-established one-parameter scaling theory~\cite{thouless1974electrons,abrahams1979scaling}, conventional 3D disordered systems exhibit both Anderson localization and MEs, whereas their one-dimensional (1D) and two-dimensional (2D) counterparts do not. However, unlike in randomly disordered systems, the localization-delocalization transitions and MEs can exist in 1D quasiperiodic lattices. A well-known example that exhibits the localization transition is the Aubry-Andr{\'e}-Harper (AAH) model~\cite{harper1955single,aubry1980analyticity}. The model has exact self-duality and there is no ME. It has been observed that when the AAH model is generalized and the self-duality is broken, MEs may appear in these 1D quasiperiodic systems~\cite{sarma1988mobility,biddle2009localization,biddle2010predicted,ganeshan2015nearest,danieli2015flat,deng2019one,li2017mobility,wang2020one,wang2020realization,wang2021duality,liu2022anomalous,wang2022quantum,gonccalves2022hidden,gonccalves2023critical,gonccalves2023renormalization,vu2023generic,wang2023engineering,qi2023multiple}. These results have been studied and confirmed in various experiments through the realization of quasiperiodic systems~\cite{roati2008anderson,luschen2018single,an2018engineering,an2021interactions,lin2022topological,wang2022observation,li2023observation}.

In recent years, there has been growing interest in investigating the phenomena of Anderson localization and MEs in non-Hermitian (NH) systems with disorder and quasiperiodicity~\cite{zeng2017anderson,longhi2019topological,longhi2019metal,jiang2019interplay,zeng2020topological,liu2020non,liu2020generalized,tzortzakakis2020non,huang2020anderson,schiffer2021anderson,tang2021localization,liu2021localization,liu2021exact,liu2021exact,cai2022localization,jiang2021mobility,jiang2021non,wu2021non,cai2022equivalence,cai2022localization,sarkar2022interplay,zeng2022real,jiang2023general,qi2023localization,padhan2024complete,acharya2024localization}. These NH systems, which arise due to interactions between the system and its surroundings, can be represented by Hamiltonians containing nonreciprocal hopping processes or complex potentials. As a result, these NH systems exhibit unique physical properties distinct from those observed in Hermitian systems. One such phenomenon is the non-Hermitian skin effect (NHSE)~\cite{yao2018edge,gong2018topological,lee2019anatomy,okuma2020topological,zhang2020correspondence,song2019non,yi2020non,guo2021exact,longhi2021phase,zeng2022real,longhi2022non,peng2022manipulating,mao2023non,lin2023topological,mao2024liouvillian}, which arises when the bulk eigenstates of an NH system under open boundary conditions (OBC) become exponentially localized near its boundaries. Another example is the NH extension of the AAH model by the complexion of the quasiperiodic potential\cite{xu2020dynamical,xu2021dynamical}. It has been shown that these localization  transitions exhibit topological characteristics and are characterized by the winding number of the energy spectrum. Meanwhile, the concept of the ME has also been extended to NH quasiperiodic systems\cite{lin2022topological,liu2020non,liu2020generalized,zeng2020winding,zhou2022driving,xia2022exact,xu2022exact,yuce2022coexistence,xu2021non,zhou2022topological,peng2023power}.

Although localization transitions and exact MEs have been primarily studied in 1D NH quasiperiodic lattices, 2D systems and materials are more prevalent, and 2D is considered the marginal dimension for localization\cite{abrahams1979scaling,fastenrath1990evidence,white2020observation}. Therefore, the study of 2D Anderson localization and MEs is of great importance for both fundamental physics and potential applications. Despite this, there is a lack of understanding regarding localization transitions and MEs in 2D NH quasiperiodic systems\cite{xu2022exact,bordia2017probing,gautier2021strongly,szabo2020mixed,vstrkalj2022coexistence,wang2023two}. Additionally, 2D disorder models with  exact MEs are rare, and the physical mechanisms that induce MEs in 2D NH systems remain obscure. As a result, analytical results of Anderson localization or MEs are especially important to study the localization phenomena of 2D NH systems.

In this work, we propose a 2D dissipative Lieb lattice (DLL) model, where imaginary quasiperiodic potentials are integrated into the Lieb lattice, which consists of equally spaced sites, and only acts on the vertices. We derive the exact expressions of the ME through analytical mapping of the 2D DLL model to the 2D isotropic AAH model. We present numerical evidence of the existence of the ME through the calculation of the fractal dimension and spatial distribution of typical eigenstates. Our analysis also shows that the introduction of imaginary quasiperiodic potentials does not modify the flat bands by examining the energy spectrum statistics. Finally, we propose a theoretical approach for implementing the DLL model using the Lindblad master equation.

The rest of the paper is organized as follows: In Sec.~\ref{section2} we introduce a 2D DLL model with imaginary quasiperiodic potentials. We obtain the exact ME analytically by mapping this model to the 2D isotropic AAH model and verify  the ME through numerical calculations of the fractal dimension and spatial distributions of two typical eigenstates in Sec.~\ref{section3}. In Sec.~\ref{section4}, we further investigate the robustness of flat bands in the presence of dissipation. Then, in Sec.~\ref{section5} we propose a theoretical scheme for realizing our model. Finally, we make a brief summary in Sec.~\ref{section6}.

\section{Model Hamiltonian}\label{section2}
\begin{figure}[t]
	\centering
	\includegraphics[width=4.5in]{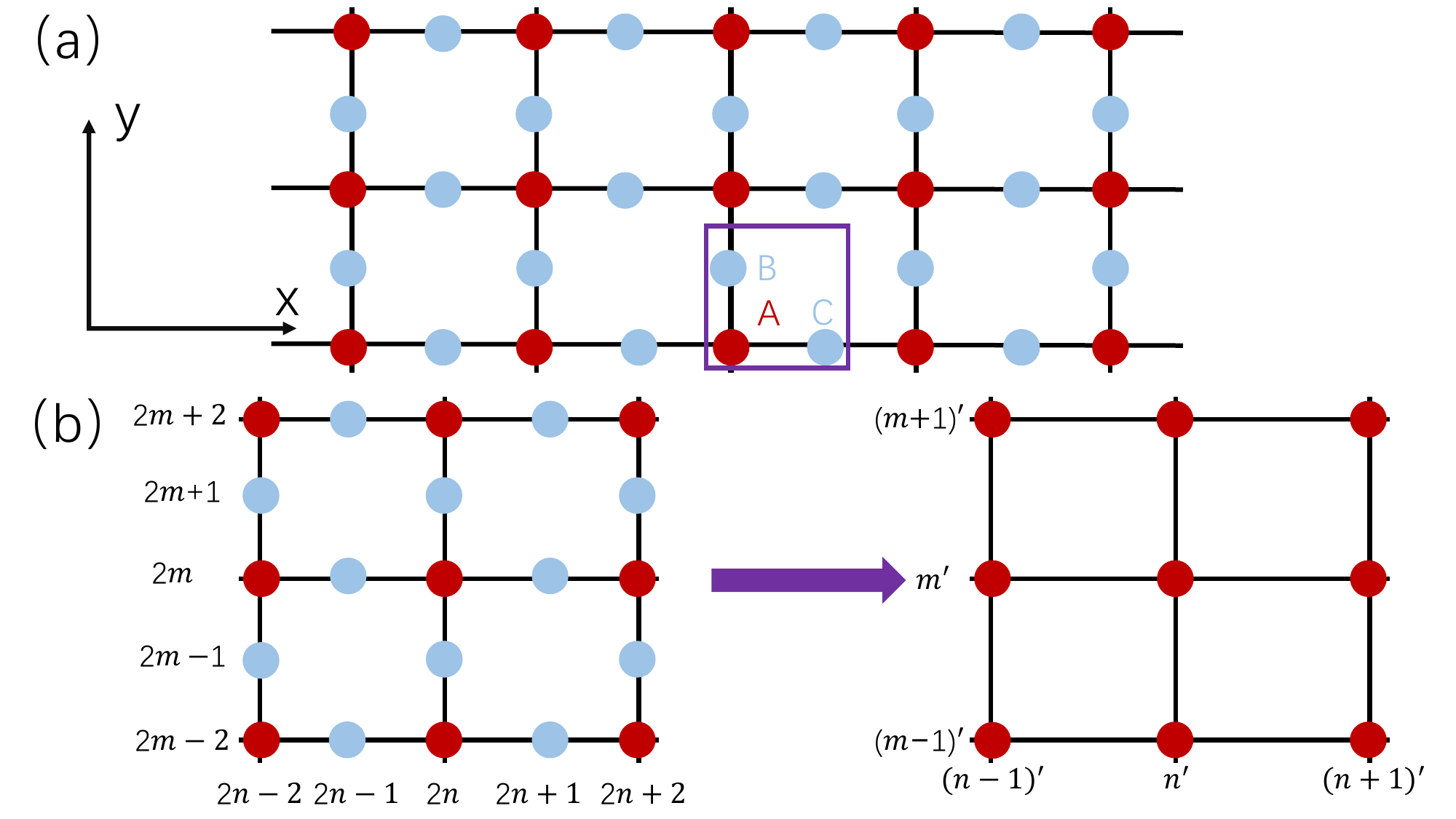}
	\caption{(Color online) Schematic illustration of the 2D dissipative Lieb lattice an d the mapping. (a) The dissipative Lieb lattice representing the edge-centered square lattice with three lattice sites [A, B, and C] per unit cell. (b) Mapping our model to the 2D non-Hermitian AAH model. Here the quasiperiodic potentials only act on the vertices (red spheres).}
	\label{fig1}
\end{figure}

We consider a 2D DLL model with imaginary quasiperiodic potentials, which is given by the following  Hamiltonian:
\begin{equation}\label{equation1}
H=\sum_{\langle kl;k'l'\rangle}t(c^{\dagger}_{kl}c_{k'l'}+h.c.)+\sum_{kl}V_{kl}c^{\dagger}_{kl}c_{kl},
\end{equation}
with 
\begin{equation}\label{equation2}
V_{kl}=\left\{ \eqalign{
2i\lambda[\cos(2\pi\alpha_x k+\theta_x)+\cos(2\pi\alpha_y l+\theta_y)], \quad & (k,l)=(2m,2n),
\cr
0, \quad & \textrm{otherwise},}
\right.
\end{equation}
where $c^{\dagger}_{kl}$ ($c_{kl}$) is the creation (annihilation) operator of a spinless fermion that acts on site $(k,l)$, $t$ represents the nearest-neighbor hopping energy, and $\lambda$ is the imaginary quasiperiodic potential strength. In this paper, without loss of generality, we set the incommensurate numbers $\alpha_x=\frac{\sqrt{5}-1}{2}$, $\alpha_y=\frac{\sqrt{2}}{2}$, the phase shifts $\theta_x=\theta_y=0$, and take periodic boundary conditions (PBC) for numerical calculations unless otherwise stated. The symbols $m$ and $n$ represent the $m$th and $n$th unit cells in the $x$ and $y$ directions, respectively. For convenience, we set $m$ ($n$)  with $L_x$ ($L_y$) being the cell number in the $x$ ($y$) direction in the absence of quasiperiodic potentials, and the indices $k$ and $l$ start from $0$ to ensure that the quasiperiodic potentials only act on the vertices [the red spheres in Fig.~\ref{fig1}(a). The Lieb lattice is a well-known member of the family of flat-band models\cite{lieb1989two}, which feature three lattice sites per unit cell [the purple square in Fig.~\ref{fig1}(a)]. This particular model has been extensively employed to investigate a wide range of fascinating physical phenomena and has also been realized in numerous systems\cite{mukherjee2015observation,vicencio2015observation,diebel2016conical,xia2018unconventional,taie2015coherent,baboux2016bosonic,slot2017experimental}. In this work, we aim to investigate the exact ME based on the 2D flat-band model with imaginary quasiperiodic potentials.

\begin{figure}[t]
	\centering
	\includegraphics[width=5.0in]{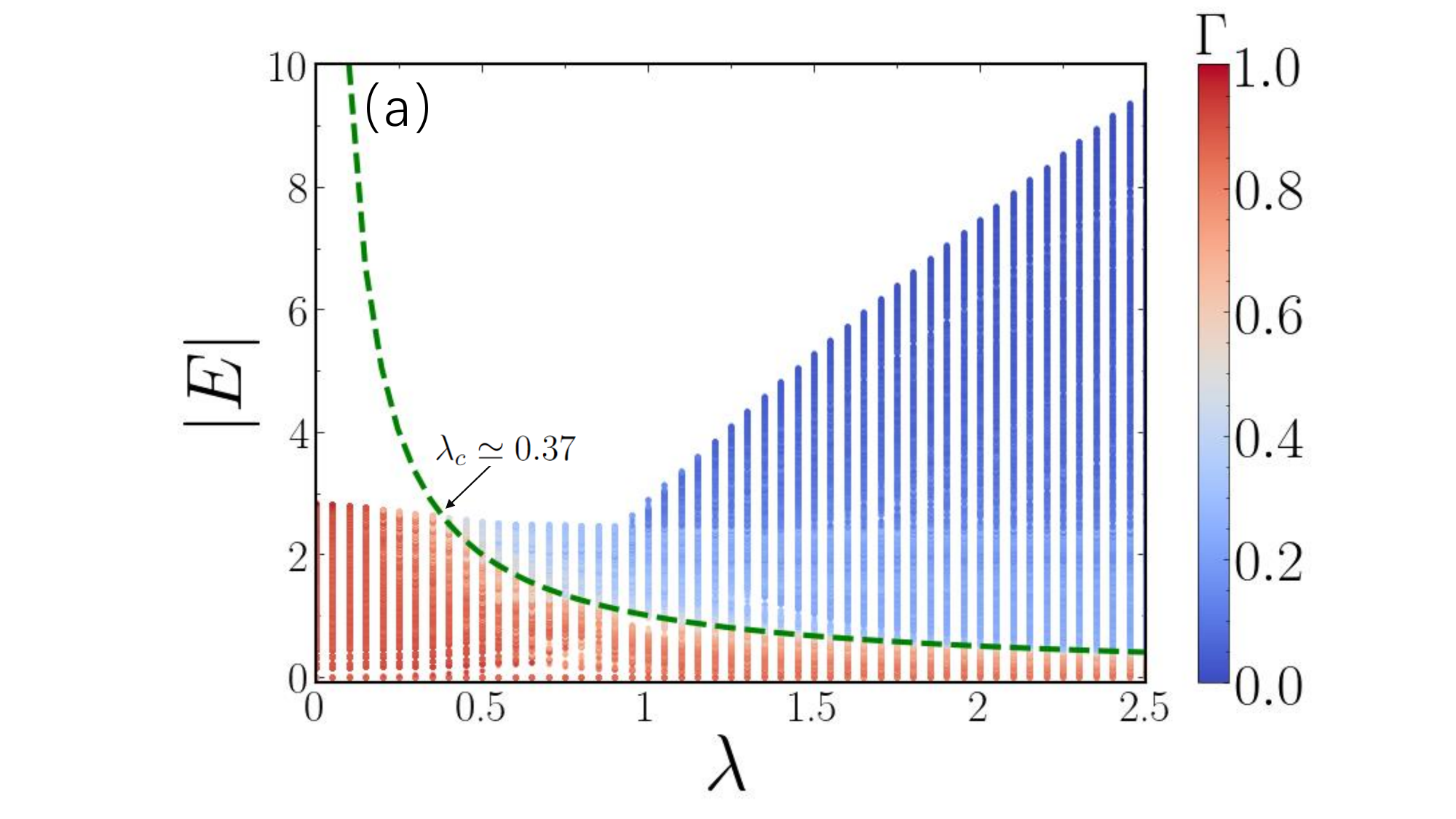}
	\includegraphics[width=5.0in]{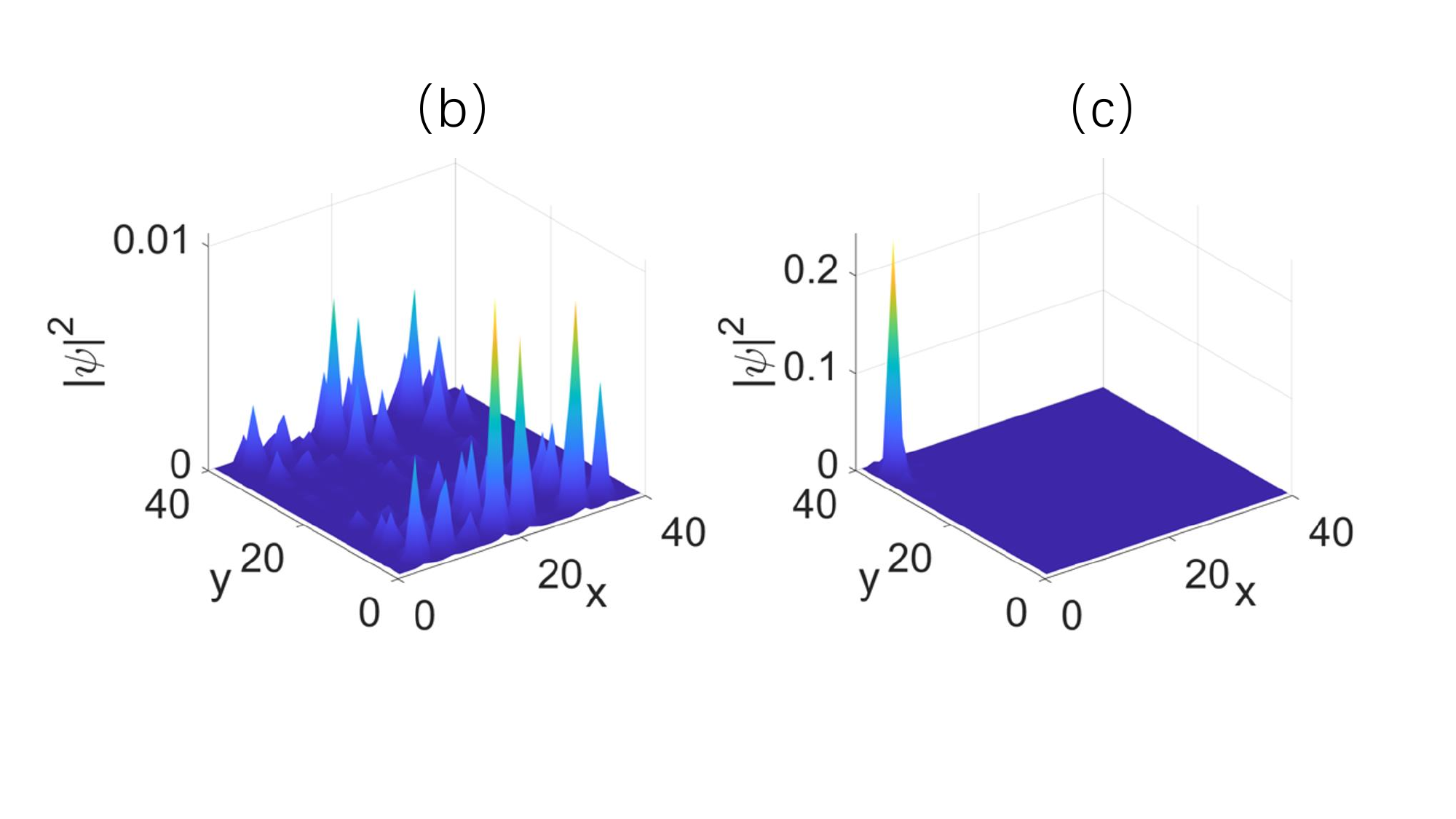}
	\vspace{-1.5cm}
	\caption{(Color online) Schematic illustration of the 2D dissipative Lieb lattice an d the mapping. (a) the dissipative Lieb lattice representing the edge-centered square lattice with three lattice sites [A, B, and C] per unit cell. The imaginary quasiperiodic potentials only act on the vertices (red spheres).  Spatial distributions of two eigenstates with the corresponding absolute value of eigenvalues (b) $|E| = 0.71$ and (c) $|E| = 1.32$, which are respectively below and above the ME ($|E_c| = 1.0$) of the DLL model with $\lambda = 1.0$ . Here, we set $t = 1$ and the lattice size $L_x = L_y = 40$.}
	\label{fig2}
\end{figure}
\section{Mapping and exact non-Hermitian mobility edge}\label{section3}
The exact MEs and localization length can be obtained by deforming the 2D DLL model to the isotropic 2D AAH model with the effective imaginary quasiperiodic potentials [as shown in Fig.~\ref{fig1}(b)]. For the Hamiltonian (\ref{equation1}), if we suppose a single-particle eigenstate is denoted as $|\Psi\rangle=\sum_{k,l}\psi_{k,l}c^{\dagger}_{k,l}|0\rangle$, then the eigenvalue equation $H|\Psi\rangle=E|\Psi\rangle$ is given by the following 
equation:
\begin{equation}\label{Lambda6}
    E\psi_{2m,2n}= t(\psi_{2m+1,2n}+\psi_{2m-1,2n}+\psi_{2m,2n+1}+\psi_{2m,2n-1})+V_{2m,2n}\psi_{2m,2n}.
\end{equation}
and
\begin{eqnarray}\label{hamS2}
E\psi_{2m+1,2n} &=& \psi_{2m+2,2n}+\psi_{2m,2n},\qquad E\psi_{2m-1,2n} = \psi_{2m,2n}+\psi_{2m-2,2n},\nonumber\\
E\psi_{2m,2n+1} &=& \psi_{2m,2n+2}+\psi_{2m,2n},\qquad E\psi_{2m,2n-1} = \psi_{2m,2n}+\psi_{2m,2n-2},
\end{eqnarray}
Base on the Eq.~(\ref{hamS2}), $\psi_{2m\pm 1,2n}$ and $\psi_{2m,2n\pm 1}$ can be replaced by $\psi_{2(m\pm 1),2n}$ and $\psi_{2m,(2n\pm 1)}$, respectively. Then, the Eq.~(\ref{Lambda6}) becomes
\begin{eqnarray}\label{eq5}
(E^2-4)\psi_{2m,2n} &=& \psi_{2(m+1),2n}+\psi_{2(m-1),2n}+\psi_{2m,2(n+1)}
+ \psi_{2m,2(n-1)} \nonumber\\
&+& EV_{2m,2n}\psi_{2m,2n}.
\end{eqnarray}
We note that the indexes are divided by $2$, i.e., 
\begin{eqnarray}
	&2m \rightarrow m', \qquad\qquad   &2(m\pm1)\rightarrow (m\pm1)', \nonumber \\
	&2n \rightarrow n', \qquad\qquad   &2(n\pm1)\rightarrow (n\pm1)', 
\end{eqnarray}
as shown in Fig.~\ref{fig1}(b), and we set the renormalized parameters 
\begin{eqnarray}
	E^2-4 \rightarrow E', \qquad\qquad  &EV \rightarrow V'.
\end{eqnarray}
Then the eigenvalue equation Eq.~(\ref{eq5}) map to the 2D isotropic AAH model with the effective imaginary quasiperiodic potentials $V'_{m',n'}=2iV'[\cos(2\pi\alpha_x m')+\cos(2\pi\alpha_y n')]$, where $V'=\lambda E$. We can analytically obtain the transition point by using the dual transformation as shown in ~\ref{Appendix_A}, and the localization-delocalization transition point is at
$|V'|=1\rightarrow|\lambda E|=1$.  Based on the previous discourse, it can be inferred that there exists a ME within our DLL model, which is given by
\begin{equation}\label{eq7}
      \centering
      |E_c|=\frac{1}{\lambda}.
\end{equation}
Following the analytical findings, the eigenstates exhibit either localized or extended behavior based on the absolute eigenvalues satisfying the conditions $|E_c|<\frac{1}{\lambda}$ or $|E_c|>\frac{1}{\lambda}$. The numerical verification of these results can be accomplished by calculating the fractal dimension, which is defined as $\Gamma=-\lim_{N\rightarrow\infty}\ln({\rm{IPR}})/\ln N$, where ${\rm{IPR}}=\sum_{k,l}|\psi_{k,l}|^4$ is the inverse participation ratio and $N=3 L_x L_y$ is the number of the total lattice sites. In the thermodynamic limit, the fractal dimension  $\Gamma$ approaches  $0$  for  extended states and $1$ for localized states. In Fig.~\ref{fig2}(a), we present $\Gamma$ of different eigenstates as the function of $\lambda$ and the corresponding absolute eigenvalues $|E|$. Our results reveal that the eigenstates within the red and blue regions exhibit localized and extended character, respectively. These two types of eigenstates are separated by a green dashed line, which is the exact ME described by Eq.~(\ref{eq7}). As indicated by the analytical findings, the value of the $\Gamma$ undergoes a drastic alteration as the energies surpass this dashed line. It has been observed that for values of the quasiperiodic potential strength $\lambda_c \lesssim 0.37$, the eigenstates exhibit a spatially extended nature. Conversely, when $\lambda_c > 0.37$, the eigenstates at high energies exhibit a localization behavior and a ME emerges. Furthermore, the ME can also be confirmed by computing the spatial distributions of eigenstates. When the value of  $\lambda $ is set to 1,  the extended eigenstate with eigenvalue $|E|=0.71$ and and the localized eigenstate with eigenvalue $|E|=1.32$ are respectively illustrated in Fig.~\ref{fig2}(b) and Fig.~\ref{fig2}(c).

By utilizing the aforementioned mapping, the localization length of the DLL model can be obtained. It is widely acknowledged that the localization length of an 2D AAH model is given by $\xi=1/\ln(|V'/t|)$. Thus, the localization length of the DLL model in both the $x$ and $y$ directions can be expressed as:
\begin{equation}\label{eq8}
\xi(E)=\frac{2}{\ln|\lambda E|}.
\end{equation}
In this context, the presence of the factor of two in the numerator arises from the fact that the system size is doubled when mapping the 2D AAH model onto the DLL model that was considered. We can define a critical exponent $\nu$ by $\xi\sim|\lambda-\lambda_c|^{-\nu}$ and determine $\nu=1$ according to Eq.~(\ref{eq8}). This result corresponds to the localization-delocalization transition observed in the isotropic 2D AAH model.

\begin{figure}[t]
	\centering
	\includegraphics[width=6.0in]{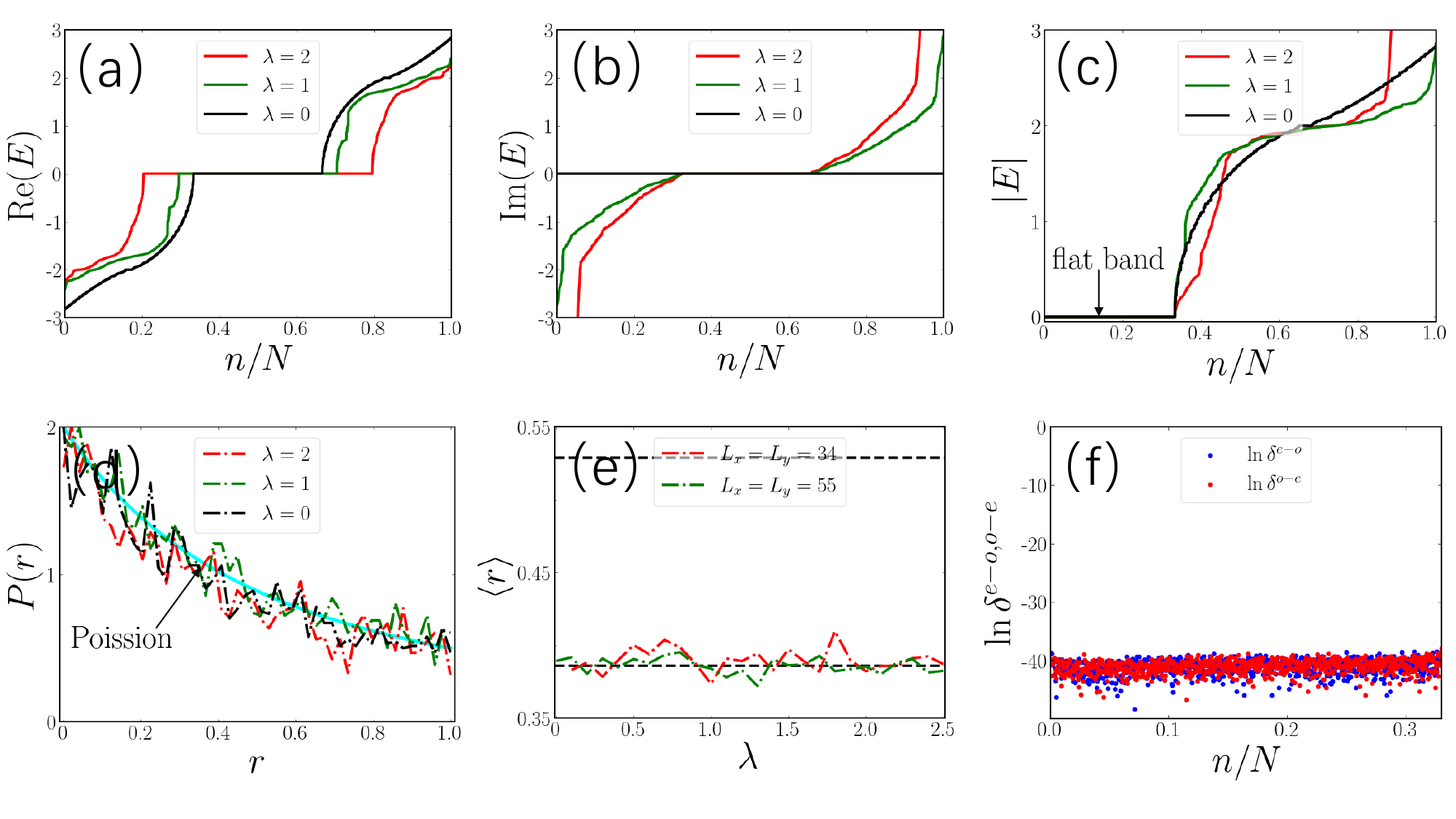}
	\caption{(Color online) Eigenvalues as a function of $n/N$ with $n$ being the index of sorted eigenenergies for the flat band, (a) the real part of eigenenergies case, (b) the imaginary part of eigenenergies case, and (c) the absolute value of eigenenergies case. (d) The probability distribution $P(r)$ of the level-spacing ratio $r$ for quasiperiodic potential strengths $\lambda=0,1.0,2.0$. (e) $\langle r\rangle$ as the function of $\lambda$ for system sizes $L_x(=L_y)=34,55$, respectively. (f) The even-odd (odd-even) level spacings of the absolute eigenvalues $\delta^{e-o}$ ($\delta^{o-e}$) for quasiperiodic potential strength $\lambda=1.0$ and systems size $L_x(=L_y)=55$.}
	\label{fig3}
\end{figure}
\section{The robustness of flat bands under dissipation}\label{section4}
In this section, we investigate the stability of flat bands, which are known to be fragile and susceptible to destruction by disorder or quasiperiodic potentials\cite{goda2006inverse,nishino2007flat,chalker2010anderson,bodyfelt2014flatbands,leykam2017localization,roy2020interplay,ahmed2022flat,lee2023critical}. In our model (\ref{equation1}), we find that these flat bands are located in the extended regions,  indicating that they may not be influenced by the Anderson localization caused by the imaginary quasiperiodic potentials. The data presented in Figs.~\ref{fig3}(a)-(c) demonstrate the flat band for the real part, imaginary part, and the absolute value of the eigenvalues, respectively. It can be observed that the flat band for the absolute eigenvalue,$|E|=0$, is unaffected by the imaginary quasiperiodic potential. As the strength of the imaginary quasiperiodic potential $\lambda$ is increased, the quantity of eigenstates found at the ${\rm{Re}}(E)=0$ increases, while the number of eigenstates in the imaginary component and the absolute value of the flat band remain unchanged. Thus, the results suggest that the existence of the flat bands within our DLL model is not affected by the introduction of imaginary quasiperiodic potentials applied to the vertex sites.

We then focus on investigating the localized eigenstates within the flat bands, namely the compact localized states. To study the influence of imaginary quasiperiodic potentials, we consider the statistical behaviors of the absolute energy levels in the flat bands by calculating the level-spacing ratio\cite{shklovskii1993statistics,oganesyan2007localization}
\begin{equation}\label{eq9}
r_n=\frac{{\rm min}(\delta E_n,\delta E_{n+1})}{{\rm max}(\delta E_n,\delta E_{n+1})}, \qquad \langle r\rangle =\frac{1}{D-1}\sum_{n=1}^{D} r_n.
\end{equation}
Here $\delta E_n=|E_{n+1}|-|E_n|$ is the absolute energy spacing, with the absolute value of eigenvalues $|E_n|$ arranged in ascending order and $n\in(0,\frac{N}{3})$ for our case. $D$ is the total number of the level-spacing ratio $r$. In the localized region, the spectral statistics exhibit characteristics of a Poisson distribution, resulting in a specific probability distribution for the level-spacing ratio $r$: $P(r)=\frac{2}{(1+r)^2}$ . This probability distribution has a mean value of $\langle r\rangle \simeq 0.387$. In contrast, in the extended region,  the spectral statistics follow Gaussian orthogonal ensemble (GOE) from random matrices theory, namely $P(r)=\frac{27(r+r^2)}{4(1+r+r^2)^{5/2}}$, and its mean value $\langle r\rangle
\simeq 0.529$. Figs.~\ref{fig3}(d) and \ref{fig3}(e) display the distribution of $r$ and the averaged $r$, respectively. It is shown that the spectral statistics exhibit Poisson distribution for different quasiperiodic potential strengths $\lambda=0,1.0,2.0$ in Fig.~\ref{fig3}(d). It indicates that the eigenstates corresponding to these energies are localized states. In Fig.~\ref{fig3}(e), the mean value $\langle r\rangle$ is $0.387$ as the function of $\lambda$ for different system sizes. To more clearly distinguish between the extended, critical, and localized states, we can define the even-odd (odd-even) level spacings of the absolute eigenvalues
$\delta^{e-o}=|E_{2n}|-|E_{2n-1}|$($\delta^{o-e}=|E_{2n+1}|-|E_{2n}|$).$|E_{2n}|$ and $|E_{2n-1}|$ denote the even and odd absolute eigenvalue in ascending order of the eigenenergy spectrum, respectively\cite{deng2019one}. In the extended region, the eigenenergy spectrum exhibits nearly doubly degenerate characteristics, resulting in a vanishing value for $\delta^{e-o}$. Consequently, there is an apparent gap between $\delta^{e-o}$ and $\delta^{o-e}$. In contrast, in the localized region, both $\delta^{e-o}$ and $\delta^{o-e}$ are almost the same, leading to the disappearance of this gap. The critical region exhibits scatter-distributed behavior for both $\delta^{e-o}$ and $\delta^{o-e}$, which are different from extended and localized phases. Our numerical results depicted in Fig.~\ref{fig3}(f) suggest that these eigenstates are located within the localized region. These quantities collectively confirm the stability of flat bands under the introduction of imaginary quasiperiodic potentials.
\section{Experimental realization}\label{section5}
The 2D Lieb lattice has been realized in numerous experiments, rendering our theoretical proposed DLL model highly feasible and capable of being simulated across diverse systems. To illustrate this, we present a theoretical framework for implementing our model, utilizing the Lindblad master equation\cite{lindblad1976generators,gorini1976completely,dalibard1992wave,carmichael1993quantum,ashida2020non}. An open quantum system consists of a quantum system that is coupled to environmental degrees of freedom. It is commonly modeled using a Markovian approximation, in which the time-evolution of the system is described by the Lindblad master equation. This equation is defined by the Liouvillian superoperator and provides a general description of the dynamics of the open quantum system. In some cases, it may be possible to describe the dynamics of an open quantum system using an effective NH Hamiltonian. This Hamiltonian takes into account the effects of the environment on the system and provides a more accurate representation of the system's behavior. However, it should be noted that not all open quantum systems can be accurately described using an effective NH Hamiltonian, and alternative methods may need to be employed in these cases.

We consider an open quantum system consisting of fermions interacting with a lossy environment, which is described by the following Lindblad master equation\cite{lindblad1976generators,li2022engineering},
\begin{eqnarray}\label{eq:lind}
\frac{d\rho}{dt}&=&-i[H_0,\rho]-\sum_{\mu}(L_\mu^\dag L_\mu\rho+\rho L^\dag_\mu L_\mu-2L_\mu\rho L_\mu^{\dagger}) \nonumber \\
&=&\mathcal{L}\rho, 
\end{eqnarray}
where $\mathcal{L}$ is named Liouvillian superoperator and $\rho$ is the density matrix. The Hamiltonian $H_0$ is defined as the sum of quadratic terms $\sum_{i,j}^{n}h_{ij}c^\dagger_i c_j$, where $h_{ij}$ are the matrix elements of $H_0$, and $c_i$ ($c_i^\dag$) is the fermionic annihilation (creation) operator of the $i$th mode. The linear quantum jump operators are given by $L_{\mu}=\sum_{i=1}^{n}l_{\mu,i}c_i$, where $l_{\mu,i}$ represents the corresponding mode-selective loss rate. From a quantum trajectory perspective, the dynamics at short times are governed by the NH Hamiltonian $H_{\mathrm{eff}}=H_0-i\sum_{\mu}L_\mu^{\dagger}L_\mu$ in the absence of quantum jump terms $L_\mu\rho L_\mu^{\dagger}$. In contrast, the long-time dynamics of the system are described by the Liouvillian superoperator $\mathcal{L}$, which the steady-state density matrix $\rho_s$ satisfies the equation $\mathcal{L}\rho_s=0$.

To establish our DLL model, we begin by considering a tight-binding Lieb lattice model that incorporates local quasiperiodic loss, with the clean Lieb lattice Hamiltonian
\begin{eqnarray}
 H_0= \sum_{\langle kl;k'l'\rangle}t(c^{\dagger}_{kl}c_{k'l'}+h.c.),
\end{eqnarray}
and the jump operators
\begin{eqnarray}\label{eq:L1}
L_{kl}=& \Gamma_{kl}c_{kl}, \quad (k,l)=(2m,2n).
\end{eqnarray}
Here, $t$ represents the nearest-neighbor hopping strength, while
$c^{\dagger}_{kl}$ ($c_{kl}$) denotes the creation (annihilation) operator of a spinless fermion that acts on site $(k,l)$. The local loss rate $\Gamma_{kl}$
is given by the formula $\Gamma_{kl}= \sqrt{2\lambda[\cos(2\pi\alpha_x k)+\cos(2\pi\alpha_y l)]}$, where $\alpha_x=\frac{\sqrt{5}-1}{2}$, $\alpha_y=\frac{\sqrt{2}}{2}$, and $\lambda$ denotes the strength of the quasiperiodic dissipation. The corresponding NH Hamiltonian is a 2D Lieb lattice model with quasiperiodic on-site loss, which can be written as
\begin{eqnarray}\label{eq:H1}
H_{\mathrm{eff}}=\sum_{\langle kl;k'l'\rangle}t(c^{\dagger}_{kl}c_{k'l'}+h.c.)-\sum_{kl}2i\lambda[\cos(2\pi\alpha_x k)+\cos(2\pi\alpha_y l)]c^{\dagger}_{kl}c_{kl}.
\end{eqnarray}
We can see that, the effective Hamiltonian $H_{\mathrm{eff}}$ is equivalent to our model Hamiltonian (\ref{equation1}), as it can be obtained by taking the Hermitian conjugate of the original Hamiltonian. Thus, it is possible to construct our DLL model in open quantum systems by introducing quasiperiodic dissipation processes alone.

\section{Summary}\label{section6}
In summary, we have investigated the exact MEs and flat bands of a DLL model with imaginary quasiperiodic potentials applied only to the vertices. We have demonstrated that the model exhibits exact MEs. The presence of the ME has been numerically confirmed by calculating the fractal dimension. Importantly, we have observed that the introduction of imaginary quasiperiodic potentials does not affect the flat bands in the system. Additionally, we propose a theoretical framework for realizing our model using the Lindblad master equation. Our findings contribute to the study of exact MEs and flat bands in 2D dissipative quasiperiodic systems.

\section*{Acknowledgments}
We thank Yucheng Wang for insightful discussions and generous help.
This work is supported by the China Postdoctoral Science Foundation (No.~2023M743267), the Key Research Projects of Zhejiang Lab (Nos.~2021PB0AC01 and 2021PB0AC02), and the National Science Foundation of China (Grant No.~12204432 and No.~62301505).

\appendix
\section*{Appendix}
\setcounter{equation}{0} \setcounter{figure}{0} \setcounter{table}{0}

\section{The 2D isotropic AAH model and localization transition}\label{Appendix_A}
We consider a 2D isotropic AAH model with a imaginary quasiperiodic potential, the Hamiltonian is described as 
\begin{equation}\label{Appendix_eq1}
H_{2D}= t\sum_{m,n}(c^{\dagger}_{m,n}c_{m+1,n}+c^{\dagger}_{m,n}c_{m,n+1}+h.c.)+2i\lambda[\cos(2\pi\alpha_x m)+\cos(2\pi\alpha_y n)]c^{\dagger}_{m,n}c_{m,n}.
\end{equation}
Suppose that an eigenstate is given by $|\Psi\rangle=\sum_{m,n}\psi_{m,n}c^{\dagger}_{m,n}|0\rangle$ and base on the Eq.~(\ref{Appendix_eq1}), we get the eigenvalue equation, i.e.:
\begin{equation}\label{Appendix_eq2}
E\psi_{m,n}= t(\psi_{m+1,n}+\psi_{m-1,n}+\psi_{m,n+1}+\psi_{m,n-1})+2i\lambda[\cos(2\pi\alpha_x m)+\cos(2\pi\alpha_y n)]\psi_{m,n}.
\end{equation}
By using the dual transformation
$\psi_{m,n}=\frac{1}{L}\sum_{\overline{m},\overline{n}}\overline{\psi}_{\overline{m},\overline{n}}e^{-i(2\pi m \alpha_x\overline{m} +2\pi n \alpha_y \overline{n})}$,
then the Eq.~(\ref{Appendix_eq2}) denotes as 
\begin{equation}\label{Appendix_eq3}
E \overline{\psi}_{\overline{m},\overline{n}}= \lambda(\overline{\psi}_{\overline{m}+1,\overline{n}}+\overline{\psi}_{\overline{m}-1,\overline{n}}+\overline{\psi}_{\overline{m},\overline{n}+1}+\overline{\psi}_{\overline{m},\overline{n}-1})+2it[\cos(2\pi\alpha_x \overline{m})+\cos(2\pi\alpha_y \overline{n})]\overline{\psi}_{\overline{m},\overline{n}}.
\end{equation}
From the above Eq.~(\ref{Appendix_eq2}) and Eq.~(\ref{Appendix_eq3}), it is invariance and self-dual $\lambda \leftrightarrow t$ when we exchange the parameter $t$ and $\lambda$. Thus, the localization-delocalization transition point is at $|\lambda/t|=1$, and no MEs exist.

\section{The mechanism of robust flat bands}
In this Appendix, we discuss the underlying mechanism for the existence of robust flat bands in our DLL model.  When the imaginary quasiperiodic potentials are absent, i.e., $V_{kl}=0$ in Eq.~(\ref{equation1}), , we transform the Hamiltonian into momentum space, which consists of a $3\times 3$ matrix due to the three inequivalent lattice sites per unit cell in Fig.~\ref{fig1}(a). This transformed Hamiltonian is then described by
\begin{equation}
H=\sum_{p}(c^{\dagger}_{p,A}, c^{\dagger}_{p,B}, c^{\dagger}_{p,C})\mathcal{H}(p)\left(
\begin{array}{c}
  c_{p,A} \\
  c_{p,B} \\
  c_{p,C} \\
\end{array}
\right).
\end{equation}
Here
\begin{equation}
\mathcal{H}(p)=\left(
\begin{array}{ccc}
  0& 2t\cos(p_y)& 2t\cos(p_x) \\
  2t\cos(p_y) & 0 & 0 \\
  2t\cos(p_x) & 0 & 0
\end{array}
\right),
\end{equation}
which has the simple form
\begin{equation}\label{hkSk}
\mathcal{H}(p)=\left(
\begin{array}{cc}
  0& S_{p} \\
  S^{\dagger}_{p} & 0 \\
\end{array}
\right).
\end{equation}
$S_{p}$ is a $1\times 2$ matrix, and we can express the singular value decomposition  of $S_{p}$ as
\begin{equation}\label{SVD1}
S_{p}=V_p\Sigma_p D_p=\sum^{r_p}_{\alpha=1}\epsilon_{p,\alpha}\phi_{p,\alpha}\psi^{\dagger}_{p,\alpha},
\end{equation}
where $V_p$ and $D_p$ are $1\times1$ and $2\times2$ unitary matrices, respectively, with their columns forming the eigenvectors of  $S_pS^{\dagger}_{p}$ and $S^{\dagger}_{p}S_p$,respectively. The rank of $S_p$, $r_p$, is equal to $1$ due to the fact  that $S_{p}$ is a $1\times 2$ matrix. The singular value matrix $\Sigma_p$ is a $1\times2$ matrix with the first column consisting of the singular value $\epsilon_{p}$ of $S_p$ and the second column being $0$. The $\alpha-$th column of $V_p$ and $D_p$ is denoted by $\phi_{p,\alpha}$ and $\psi_{p,\alpha}$,respectively, where being the $\alpha-$th left (right) singular eigenvector of $S_p$. Because $\phi_p=V_p$ and $\psi_p$ is the first column of $D_p$, $\alpha$ has to equal to $1$. Using Eq.~(\ref{SVD1}), we perform a unitary transformation on the $\mathcal{H}(p)$
\begin{equation}
\mathcal{H}(p)=\left(
\begin{array}{cc}
  V_p& 0 \\
  0 & D_p \\
\end{array}
\right)\left(
\begin{array}{cc}
  0 & \Sigma_p \\
  \Sigma^{T}_p & 0 \\
\end{array}
\right)\left(
\begin{array}{cc}
  V^{\dagger}_p& 0 \\
  0 & D^{\dagger}_p \\
\end{array}
\right).
\end{equation}
Due to the fact that $\Sigma_p=(\epsilon_{p}, 0)$, $\mathcal{H}(p)$ is similar to a matrix containing one zero row and column, which implies that $\mathcal{H}(p)$ contains at least one zero mode for any $p$. As a result, the Lieb lattice possesses one flat band that is pinned at zero energy.

Based on the above discussions, when the matrix $S_p$ in Eq.~(\ref{hkSk}) is a $N_1\times N_2$ matrix, $\mathcal{H}(p)$ necessarily possesses at least $|N_1-N_2|$ zero modes at at any given momentum value, indicating that this system exhibits at least $|N_1-N_2|$ flat bands. This can also be clearly seen from the real space. When imaginary quasiperiodic potentials are added on vertices of the Lieb lattice, $S_p$ remains unaffected, and hence the flat band is robust. However, when these potentials are introduced on the edges or facilitate hopping between the lattice sites B and C, $\mathcal{H}(p)$ can not be written as Eq.~(\ref{hkSk}), leading to the destruction of the flat bands.


\section*{References}
\begin{thebibliography}{100}

\bibitem{anderson1958absence}{P. W. Anderson, Absence of Diffusion in Certain Random Lattices, \color{blue} \href {https://doi.org/10.1103/PhysRev.109.1492} {\it Phys. Rev.} \textbf{109}, 1492 (1958).}

\bibitem{lee1985disordered}{P. A. Lee and T. V. Ramakrishnan, Disordered electronic systems, \color{blue} \href {https://doi.org/10.1103/RevModPhys.57.287} {\it Rev. Mod. Phys.} \textbf{57}, 287 (1985).}

\bibitem{kramer1993localization}{B. Kramer and A. MacKinnon, Localization: Theory and experiment, \color{blue} \href {https://dx.doi.org/10.1088/0034-4885/56/12/001} {\it Rep. Prog. Phys.} \textbf{56}, 1469 (1993).}

\bibitem{abrahams201050}{E. Abrahams, 50 Years of Anderson Localization, (world scientific, 2010)}

\bibitem{evers2008anderson}{F. Evers and A. D. Mirlin, Anderson Transitions, \color{blue} \href {https://doi.org/10.1103/RevModPhys.80.1355} {\it Rev. Mod. Phys.} \textbf{80}, 1355 (2008).}

\bibitem{thouless1974electrons}{D. J. Thouless, Electrons in disordered systems and the theory of localization, \color{blue} \href {https://doi.org/10.1016/0370-1573(74)90029-5} {\it Phys. Rep.} \textbf{13}, 93 (1974).}

\bibitem{abrahams1979scaling}{E. Abrahams, P. W. Anderson, D. C. Licciardello, and T. V. Ramakrishnan, Scaling Theory of Localization: Absence of Quantum Diffusion in Two Dimensions, \color{blue} \href {https://link.aps.org/doi/10.1103/PhysRevLett.42.673} {\it Phys. Rev. Lett.} \textbf{42}, 673 (1979).}

\bibitem{harper1955single}{P. G. Harper, Single band motion of conduction electrons
in a uniform magnetic field, \color{blue} \href {https://dx.doi.org/10.1088/0370-1298/68/10/304} {\it Proc. Phys. Soc., Sect A} \textbf{68}, 874 (1955).}

\bibitem{aubry1980analyticity}{S. Aubry and G. Andr{\'e}, Analyticity breaking and Anderson localization in incommensurate lattices, \color{blue} \href {https://chaos.if.uj.edu.pl/~delande/Lectures/files/An.Is.Phys.Soc.pdf} {\it Ann. Israel
Phys. Soc.} \textbf{3}, 133 (1980).}

\bibitem{sarma1988mobility}{S. D. Sarma, S. He, and X. C. Xie, Mobility Edge in a Model One-Dimensional Potential, \color{blue} \href {https://doi.org/10.1103/PhysRevLett.61.2144} {\it Phys. Rev. Lett.} \textbf{61}, 2144 (1988).}

\bibitem{biddle2009localization}{J. Biddle, B. Wang, D. J. Priour, Jr., and S. Das Sarma, Localization in one-dimensional incommensurate lattices beyond the Aubry-Andr{\'e} model, \color{blue} \href {https://link.aps.org/doi/10.1103/PhysRevA.80.021603} {\it Phys. Rev. A} \textbf{80}, 021603(R)  (2009).}

\bibitem{biddle2010predicted}{J. Biddle and S. Das Sarma, Predicted Mobility Edges in One-Dimensional Incommensurate Optical Lattices: An Exactly Solvable Model of Anderson Localization, \color{blue} \href {https://link.aps.org/doi/10.1103/PhysRevLett.104.070601} {\it Phys. Rev. Lett.} \textbf{104}, 070601 (2010).}

\bibitem{ganeshan2015nearest}{S. Ganeshan, J. H. Pixley, and S. Das Sarma, Nearest Neighbor Tight Binding Models with an Exact Mobility Edge in One Dimension,  \color{blue} \href {https://link.aps.org/doi/10.1103/PhysRevLett.114.146601} {\it Phys. Rev. Lett.} \textbf{114}, 146601 (2015).}

\bibitem{danieli2015flat}{C. Danieli, J. D. Bodyfelt, and S. Flach, Flat-band engineering of mobility edges,  \color{blue} \href {https://link.aps.org/doi/10.1103/PhysRevB.91.235134} {\it Phys. Rev. B} \textbf{91}, 235134 (2015).} 

\bibitem{deng2019one}{X. Deng, S. Ray, S. Sinha, G. V. Shlyapnikov, and L. Santos, One-Dimensional Quasicrystals with Power-Law Hopping,  \color{blue} \href {https://link.aps.org/doi/10.1103/PhysRevLett.123.025301} {\it Phys. Rev. Lett.} \textbf{123}, 025301 (2019).} 

\bibitem{li2017mobility}{X. Li, X. Li, and S. Das Sarma, Mobility edges in one dimensional bichromatic incommensurate potentials,  \color{blue} \href {https://link.aps.org/doi/10.1103/PhysRevB.96.085119} {\it Phys. Rev. B} \textbf{96}, 085119 (2017).}

\bibitem{wang2020one}{Y. Wang, X. Xia, L. Zhang, H. Yao, S. Chen, J. You, Q. Zhou, and X.-J. Liu, One-Dimensional Quasiperiodic Mosaic Lattice with Exact Mobility Edges,  \color{blue} \href {https://link.aps.org/doi/10.1103/PhysRevLett.125.196604} {\it Phys. Rev. Lett.} \textbf{125}, 196604 (2020).}

\bibitem{wang2020realization}{Y. Wang, L. Zhang, S. Niu, D. Yu, and X.-J. Liu, Realization and Detection of Nonergodic Critical Phases in an Optical Raman Lattice,  \color{blue} \href {https://link.aps.org/doi/10.1103/PhysRevLett.125.073204} {\it Phys. Rev. Lett.} \textbf{125}, 073204 (2020).} 

\bibitem{wang2021duality}{Y. Wang, X. Xia, Y. Wang, Z. Zheng, and X.-J. Liu, Duality between two generalized Aubry-André models with exact mobility edges,  \color{blue} \href {https://link.aps.org/doi/10.1103/PhysRevB.103.174205} {\it Phys. Rev. B} \textbf{103}, 174205 (2021).}

\bibitem{liu2022anomalous}{T. Liu, X. Xia, S. Longhi, and L. Sanchez-Palencia, Anmalous mobility edges in one-dimensional quasiperiodic models,  \color{blue} \href {https://scipost.org/10.21468/SciPostPhys.12.1.027} {\it SciPost Phys.} \textbf{12}, 027 (2022).}

\bibitem{wang2022quantum}{Y. Wang, L. Zhang,W. Sun, T.-F. J. Poon, and X.-J. Liu, Quantum phase with coexisting localized, extended, and critical zones,  \color{blue} \href {https://link.aps.org/doi/10.1103/PhysRevB.106.L140203} {\it Phys. Rev. B} \textbf{106}, L140203 (2022).}

\bibitem{gonccalves2022hidden}{M. Gonçalves, B. Amorim, E. Castro, and P. Ribeiro, Hidden Dualities in 1D Quasiperiodic Lattice Models,  \color{blue} \href {https://scipost.org/10.21468/SciPostPhys.13.3.046} {\it SciPost Phys.} \textbf{13}, 046 (2022).}

\bibitem{gonccalves2023critical}{M. Gonçalves, B. Amorim, E. Castro, and P. Ribeiro,  Critical Phase Dualities in 1D Exactly Solvable Quasiperiodic Models, \color{blue} \href {https://link.aps.org/doi/10.1103/PhysRevLett.131.186303} {\it  Phys. Rev. Lett.} \textbf{131}, 186303 (2023).}

\bibitem{gonccalves2023renormalization}{M. Gonçalves, B. Amorim, E. V. Castro, and P. Ribeiro,  Renormalization group theory of one-dimensional quasiperiodic lattice models with commensurate approximants, \color{blue} \href {https://link.aps.org/doi/10.1103/PhysRevB.108.L100201} {\it  Phys. Rev. B} \textbf{108}, L100201 (2023).}

\bibitem{vu2023generic}{D. D. Vu and S. Das Sarma, Generic mobility edges in several classes of duality-breaking one-dimensional quasiperiodic potentials, \color{blue} \href {https://link.aps.org/doi/10.1103/PhysRevB.107.224206} {\it  Phys. Rev. B} \textbf{107}, 224206 (2023).}

\bibitem{wang2023engineering}{Z. Wang, Y. Zhang, L. Wang, and S. Chen, Engineering mobility in quasiperiodic lattices with exact mobility edges, \color{blue} \href {https://link.aps.org/doi/10.1103/PhysRevB.108.174202} {\it  Phys. Rev. B} \textbf{108}, 174202 (2023).}

\bibitem{qi2023multiple}{R. Qi, J. Cao, and X.-P. Jiang, Multiple localization transitions and novel quantum phases induced by a staggered on-site potential, \color{blue} \href {https://link.aps.org/doi/10.1103/PhysRevB.107.224201} {\it  Phys. Rev. B} \textbf{107}, 224201 (2023).}

\bibitem{roati2008anderson}{G. Roati, C. D’Errico, L. Fallani, M. Fattori, C. Fort, M. Zaccanti, G. Modugno, M. Modugno, and M. Inguscio, Anderson localization of a non-interacting Bose-Einstein condensate, \color{blue} \href {https://doi.org/10.1038/nature07071} {\it   Nature (London)} \textbf{453}, 895 (2008).}

\bibitem{luschen2018single}{H. P. Lüschen, S. Scherg, T. Kohlert, M. Schreiber, P. Bordia, X. Li, S. D. Sarma, and I. Bloch, Single-Particle Mobility Edge in a One-Dimensional Quasiperiodic Optical Lattice, \color{blue} \href {https://link.aps.org/doi/10.1103/PhysRevLett.120.160404} {\it  Phys. Rev. Lett.} \textbf{120}, 160404 (2018).}

\bibitem{an2018engineering}{F. A. An, E. J. Meier, and B. Gadway, Engineering a Flux-Dependent Mobility Edge in Disordered Zigzag Chains, \color{blue} \href {https://link.aps.org/doi/10.1103/PhysRevX.8.031045} {\it   Phys. Rev. X} \textbf{8}, 031045 (2018).}

\bibitem{an2021interactions}{F. A. An, K. Padavi´c, E. J. Meier, S. Hegde, S. Ganeshan, J. H. Pixley, S. Vishveshwara, and B. Gadway, Interactions and Mobility Edges: Observing the Generalized Aubry-Andr{\'e} Model, \color{blue} \href {https://link.aps.org/doi/10.1103/PhysRevLett.126.040603} {\it   Phys. Rev. Lett.} \textbf{126}, 040603 (2021).}

\bibitem{lin2022topological}{Q. Lin, T. Li, L. Xiao, K. Wang, W. Yi, and P. Xue, Topological Phase Transitions and Mobility Edges in Non-Hermitian Quasicrystals, \color{blue} \href {https://link.aps.org/doi/10.1103/PhysRevLett.129.113601} {\it  Phys. Rev. Lett.} \textbf{129}, 113601 (2022).}

\bibitem{wang2022observation}{Y. Wang, J.-H. Zhang, Y. Li, J. Wu, W. Liu, F. Mei, Y. Hu, L. Xiao, J. Ma, C. Chin, and S. Jia, Observation of Interaction-Induced Mobility Edge in an Atomic Aubry-Andr{\'e} Wire, \color{blue} \href {https://link.aps.org/doi/10.1103/PhysRevLett.129.103401} {\it  Phys. Rev. Lett.} \textbf{129}, 103401 (2022).} 

\bibitem{li2023observation}{H. Li, Y.-Y. Wang, Y.-H. Shi, K. Huang, X. Song, G.-H. Liang, Z.-Y. Mei, B. Zhou, H. Zhang, J.-C. Zhang, S. Chen, S. Zhao, Y. Tian, Z.-Y. Yang, Z. Xiang, K. Xu, D. Zheng, and H. Fan, Observation of critical phase transition in a generalized Aubry-Andr{\'e}-Harper model with superconducting circuits.  \color{blue} \href {https://doi.org/10.1038/s41534-023-00712-w} {\it  npj Quantum Inf} \textbf{9}, 40 (2023).} 

\bibitem{zeng2017anderson}{Q.-B. Zeng, S. Chen, and R. Lü, Anderson localization in the Non-Hermitian Aubry-Andr{\'e}-Harper model with physical gain and loss, \color{blue} \href {https://link.aps.org/doi/10.1103/PhysRevA.95.062118} {\it  Phys. Rev. A} \textbf{95}, 062118 (2017).}

\bibitem{longhi2019topological}{S. Longhi, Topological Phase Transition in non-Hermitian Quasicrystals, \color{blue} \href {https://link.aps.org/doi/10.1103/PhysRevLett.122.237601} {\it  Phys. Rev. Lett.} \textbf{122}, 237601 (2019).}

\bibitem{longhi2019metal}{S. Longhi, Metal-insulator phase transition in a non-Hermitian Aubry-Andr{\'e}-Harper model, \color{blue} \href {https://link.aps.org/doi/10.1103/PhysRevB.100.125157} {\it  Phys. Rev. B} \textbf{100}, 125157 (2019).}

\bibitem{jiang2019interplay}{H. Jiang, L.-J. Lang, C. Yang, S.-L. Zhu, and S. Chen, Interplay of non-Hermitian skin effects and Anderson localization in nonreciprocal quasiperiodic lattices, \color{blue} \href {https://link.aps.org/doi/10.1103/PhysRevB.100.054301} {\it  Phys. Rev. B} \textbf{100}, 054301 (2019).}

\bibitem{zeng2020topological}{Q.-B. Zeng, Y.-B. Yang, and Y. Xu, Topological phases in non-Hermitian Aubry-André-Harper models, \color{blue} \href {https://link.aps.org/doi/10.1103/PhysRevB.100.054301} {\it  Phys. Rev. B} \textbf{101}, 020201 (2020).}

\bibitem{liu2020non}{Y. Liu, X.-P. Jiang, J. Cao, and S. Chen, Non-Hermitian mobility edges in one-dimensional quasicrystals with parity-time symmetry, \color{blue} \href {https://link.aps.org/doi/10.1103/PhysRevB.101.174205} {\it  Phys. Rev. B} \textbf{101}, 174205 (2020).}

\bibitem{liu2020generalized}{T. Liu, H. Guo, Y. Pu, and S. Longhi, Generalized Aubry-André self-duality and mobility edges in non-Hermitian quasiperiodic lattices, \color{blue} \href {https://link.aps.org/doi/10.1103/PhysRevB.102.024205} {\it  Phys. Rev. B} \textbf{102}, 024205 (2020).}

\bibitem{tzortzakakis2020non}{A. F. Tzortzakakis, K. G. Makris, and E. N. Economou, Non-Hermitian disorder in two-dimensional optical lattices, \color{blue} \href {https://link.aps.org/doi/10.1103/PhysRevB.101.014202} {\it  Phys. Rev. B} \textbf{101}, 014202 (2020).}

\bibitem{huang2020anderson}{Y. Huang and B. I. Shklovskii, Anderson transition in three-dimensional systems with non-Hermitian disorder, \color{blue} \href {https://link.aps.org/doi/10.1103/PhysRevB.101.014204} {\it  Phys. Rev. B} \textbf{101}, 014204 (2020).}

\bibitem{schiffer2021anderson}{S. Schiffer, X.-J. Liu, H. Hu, and J. Wang, Anderson localization transition in a robust PT-symmetric phase of a generalized Aubry-André model, \color{blue} \href {https://link.aps.org/doi/10.1103/PhysRevA.103.L011302} {\it  Phys. Rev. A} \textbf{103}, L011302 (2021).}

\bibitem{tang2021localization}{L.-Z. Tang, G.-Q. Zhang, L.-F. Zhang, and D.-W. Zhang, Localization and topological transitions in non-Hermitian quasiperiodic lattices, \color{blue} \href {https://link.aps.org/doi/10.1103/PhysRevA.103.033325} {\it  Phys. Rev. A} \textbf{103}, 033325 (2021).}

\bibitem{liu2021localization}{Y. Liu, Q. Zhou, and S. Chen, Localization transition, spectrum structure, and winding numbers for one-dimensional non-Hermitian quasicrystals, \color{blue} \href {https://link.aps.org/doi/10.1103/PhysRevB.104.024201} {\it  Phys. Rev. B} \textbf{104}, 024201 (2021).}
 
\bibitem{liu2021exact}{Y. Liu, Y. Wang, Z. Zheng, and S. Chen, Exact non-Hermitian mobility edges in one-dimensional quasicrystal lattice with exponentially decaying hopping and its dual lattice, \color{blue} \href {https://link.aps.org/doi/10.1103/PhysRevB.103.134208} {\it  Phys. Rev. B} \textbf{103}, 134208 (2021).}

\bibitem{liu2021exact1}{Y. Liu, Y. Wang, X.-J. Liu, Q. Zhou, and S. Chen, Exact mobility edges, PT-symmetry breaking, and skin effect in one-dimensional non-Hermitian quasicrystals, \color{blue} \href {https://link.aps.org/doi/10.1103/PhysRevB.103.014203} {\it  Phys. Rev. B} \textbf{103}, 014203 (2021).}

\bibitem{cai2022localization}{X. Cai, Localization transitions and winding numbers for non-Hermitian Aubry-André-Harper models with off-diagonal modulations, \color{blue} \href {https://link.aps.org/doi/10.1103/PhysRevB.106.214207} {\it  Phys. Rev. B} \textbf{106}, 214207 (2022).}
 
\bibitem{jiang2021mobility}{X.-P. Jiang, Y. Qiao, and J. Cao, Mobility edges and reentrant localization in one-dimensional dimerized non-Hermitian quasiperiodic lattice, \color{blue} \href {https://iopscience.iop.org/article/10.1088/1674-1056/ac11e5} {\it  Chin. Phys. B.} \textbf{30}, 097202 (2021).} 

\bibitem{jiang2021non}{X.-P. Jiang, Y. Qiao, and J. Cao, Non-Hermitian Kitaev chain with complex periodic and quasiperiodic potentials, \color{blue} \href {https://iopscience.iop.org/article/10.1088/1674-1056/abfa08} {\it  Chin. Phys. B.} \textbf{30}, 077101 (2021).}

\bibitem{wu2021non}{C. Wu, J. Fan, G. Chen, and S. Jia, Non-Hermiticity-induced reentrant localization in a quasiperiodic lattice, \color{blue} \href {https://iopscience.iop.org/article/10.1088/1367-2630/ac430b/meta} {\it  New J. Phys.} \textbf{23}, 123048 (2021).} 

\bibitem{cai2022equivalence}{X. Cai and S.-J. Jiang, Equivalence and superposition of real and imaginary quasiperiodicities, \color{blue} \href {https://iopscience.iop.org/article/10.1088/1367-2630/ac99f5} {\it  New J. Phys.} \textbf{24}, 113001 (2022).}

\bibitem{cai2022localization}{X. Cai, Localization transitions and winding numbers for non-Hermitian Aubry-André-Harper models with off-diagonal modulations, \color{blue} \href {https://link.aps.org/doi/10.1103/PhysRevB.106.214207} {\it  Phys. Rev. B} \textbf{106}, 214207  (2022).}

\bibitem{sarkar2022interplay}{R. Sarkar, S. S. Hegde, and A. Narayan, Interplay of disorder and point-gap topology: Chiral modes, localization, and non-Hermitian Anderson skin effect in one dimension, \color{blue} \href {https://link.aps.org/doi/10.1103/PhysRevB.106.014207} {\it  Phys. Rev. B} \textbf{106}, 014207 (2022).}

\bibitem{zeng2022real}{Q.-B. Zeng and R. Lü, Real spectra, Anderson localization, and topological phases in one-dimensional quasireciprocal systems, \color{blue} \href {https://iopscience.iop.org/article/10.1088/1367-2630/ac61d0} {\it  New J. Phys.} \textbf{24}, 043023 (2022).}

\bibitem{jiang2023general}{S.-L. Jiang, Y. Liu, and L.-J. Lang, General mapping of one-dimensional non-Hermitian mosaic models to non-mosaic counterparts: Mobility edges and Lyapunov exponents, \color{blue} \href {https://iopscience.iop.org/article/10.1088/1674-1056/ace426} {\it  Chin. Phys. B.} \textbf{32}, 097204 (2023).}

\bibitem{qi2023localization}{R. Qi, J. Cao, and X.-P. Jiang, Localization and mobility edges in non-Hermitian disorder-free lattices, \color{blue} \href {https://arxiv.org/abs/2306.03807} {\it  arXiv:} 2306.03807 (2023).}

\bibitem{padhan2024complete}{A. Padhan, S. R. Padhi, and T. Mishra, Complete delocalization and reentrant topological transition in a non-Hermitian quasiperiodic lattice, \color{blue} \href {https://link.aps.org/doi/10.1103/PhysRevB.109.L020203} {\it  Phys. Rev. B} \textbf{109}, L020203 (2024).}

\bibitem{acharya2024localization}{A. P. Acharya and S. Datta, Localization transitions in a non-Hermitian quasiperiodic lattice, \color{blue} \href {https://link.aps.org/doi/10.1103/PhysRevB.109.024203} {\it  Phys. Rev. B} \textbf{109}, 024203 (2024).}

\bibitem{yao2018edge}{S. Yao and Z. Wang, Edge States and Topological Invariants of Non-Hermitian Systems, \color{blue} \href {https://link.aps.org/doi/10.1103/PhysRevLett.121.086803} {\it  Phys. Rev. Lett.} \textbf{121}, 086803 (2018).}

\bibitem{gong2018topological}{Z. Gong, Y. Ashida, K. Kawabata, K. Takasan, S. Higashikawa, and M. Ueda, Topological Phases of Non-Hermitian Systems, \color{blue} \href {https://link.aps.org/doi/10.1103/PhysRevX.8.031079} {\it  Phys. Rev. X} \textbf{8}, 031079 (2018).}

\bibitem{lee2019anatomy}{C. H. Lee and R. Thomale, Anatomy of skin modes and topology in non-hermitian systems, \color{blue} \href {https://link.aps.org/doi/10.1103/PhysRevB.99.201103} {\it  Phys. Rev. B} \textbf{99}, 201103 (2019).}

\bibitem{okuma2020topological}{N. Okuma, K. Kawabata, K. Shiozaki, and M. Sato, Topological Origin of Non-Hermitian Skin Effects, \color{blue} \href {https://link.aps.org/doi/10.1103/PhysRevLett.124.086801} {\it  Phys. Rev. Lett.} \textbf{124}, 086801 (2020)}

\bibitem{zhang2020correspondence}{K. Zhang, Z. Yang, and C. Fang, Correspondence between Winding Numbers and Skin Modes in Non-Hermitian Systems, \color{blue} \href {https://link.aps.org/doi/10.1103/PhysRevLett.124.086801} {\it  Phys. Rev. Lett.} \textbf{125}, 126402 (2020).}

\bibitem{song2019non}{F. Song, S. Yao, and Z. Wang, Non-Hermitian Skin Effect and Chiral Damping in Open Quantum Systems, \color{blue} \href {https://link.aps.org/doi/10.1103/PhysRevLett.123.170401} {\it  Phys. Rev. Lett.} \textbf{123}, 170401 (2019).}

\bibitem{yi2020non}{F. Song, S. Yao, and Z. Wang, Y. Yi and Z. Yang, Non-Hermitian Skin Modes Induced by On-Site Dissipations and Chiral Tunneling Effect, \color{blue} \href {https://link.aps.org/doi/10.1103/PhysRevLett.125.186802} {\it Phys. Rev. Lett.} \textbf{125}, 186802 (2020).}

\bibitem{guo2021exact}{C.-X. Guo, C.-H. Liu, X.-M. Zhao, Y. Liu, and S. Chen,
Exact Solution of Non-Hermitian Systems with Generalized Boundary Conditions: Size-Dependent Boundary Effect and Fragility of the Skin Effect, \color{blue} \href {https://link.aps.org/doi/10.1103/PhysRevLett.127.116801} {\it  Phys. Rev. Lett.} \textbf{127}, 116801 (2021).}

\bibitem{longhi2021phase}{S. Longhi, Phase transitions in a non-Hermitian  Aubry-Andr{\'e}-Harper model, \color{blue} \href {https://link.aps.org/doi/10.1103/PhysRevB.103.054203} {\it  Phys. Rev. B} \textbf{103}, 054203 (2021).}

\bibitem{zeng2022real}{Q.-B. Zeng and R. L\"u, Real spectra and phase transition
of skin effect in nonreciprocal systems, \color{blue} \href {https://link.aps.org/doi/10.1103/PhysRevB.105.245407} {\it  Phys. Rev. B} \textbf{105}, 245407 (2022).}

\bibitem{longhi2022non}{S. Longhi, Non-Hermitian skin effect and self-acceleration, \color{blue} \href {https://link.aps.org/doi/10.1103/PhysRevB.105.245143} {\it  Phys. Rev. B} \textbf{105}, 245143 (2022).}

\bibitem{peng2022manipulating}{Y. Peng, J. Jie, D. Yu, and Y. Wang, Manipulating the non-Hermitian skin effect via electric fields, \color{blue} \href {https://link.aps.org/doi/10.1103/PhysRevB.106.L161402} {\it  Phys. Rev. B} \textbf{106}, L161402 (2022).}

\bibitem{mao2023non}{L. Mao, Y. Hao, and L. Pan, Non-Hermitian skin effect in a one-dimensional interacting Bose gas, \color{blue} \href {https://link.aps.org/doi/10.1103/PhysRevA.107.043315} {\it  Phys. Rev. A} \textbf{107}, 043315 (2023).}

\bibitem{lin2023topological}{R. Lin, T. Tai, L. Li, C. H. Lee, Topological non-Hermitian skin effect, \color{blue} \href {https://link.springer.com/article/10.1007/s11467-023-1309-z} {\it  Front. Phys.} \textbf{18}, 53605 (2023).}

\bibitem{mao2024liouvillian}{L. Mao, M. J. Tao, H. Hu, L. Pan, Liouvillian skin effect in a one-dimensional open many-body quantum system with generalized boundary conditions, \color{blue} \href {https://arxiv.org/abs/2401.15614} {\it  arXiv:} 2401.15614 (2024).}

\bibitem{xu2020dynamical}{Z. Xu, H. Huangfu, Y. Zhang, and S. Chen, Dynamical observation of mobility edges in one-dimensional incommensurate optical lattices, \color{blue} \href {https://iopscience.iop.org/article/10.1088/1367-2630/ab64b2} {\it  New J. Phys.} \textbf{22}, 013036 (2020).}

\bibitem{xu2021dynamical}{Z. Xu and S. Chen, Dynamical evolution in a one-dimensional incommensurate lattice with 
PT symmetry, \color{blue} \href {https://link.aps.org/doi/10.1103/PhysRevA.103.043325} {\it  Phys. Rev. A} \textbf{103}, 043325 (2021).}

\bibitem{zeng2020winding}{Q.-B. Zeng and Y. Xu, Winding numbers and generalized mobility edges in non-Hermitian systems, \color{blue} \href {https://link.aps.org/doi/10.1103/PhysRevResearch.2.033052} {\it  Phys. Rev. Res.} \textbf{2}, 033052 (2020).}

\bibitem{zhou2022driving}{W. Han and L. Zhou, Dimerization-induced mobility edges and multiple reentrant localization transitions in non-Hermitian quasicrystals, \color{blue} \href {https://link.aps.org/doi/10.1103/PhysRevB.105.054204} {\it  Phys. Rev. B} \textbf{105}, 054204 (2022).}

\bibitem{xia2022exact}{W. Han and L. Zhou, X. Xia, K. Huang, S. Wang, and X. Li, Exact mobility edges in the non-Hermitian $t_1-t_2$ model: Theory and possible experimental realizations, \color{blue} \href {https://link.aps.org/doi/10.1103/PhysRevB.105.014207} {\it  Phys. Rev. B} \textbf{105}, 014207 (2022).}

\bibitem{xu2022exact}{Z. Xu, X. Xia, and S. Chen, Exact mobility edges and topological phase transition in two-dimensional non-Hermitian quasicrystals, \color{blue} \href {https://link.springer.com/article/10.1007/s11433-021-1802-4} {\it  Sci. China Phys. Mech. Astron.} \textbf{65}, 227211 (2022).}

\bibitem{yuce2022coexistence}{C. Yuce and H. Ramezani, Coexistence of extended and localized states in the one-dimensional non-Hermitian Anderson model, \color{blue} \href {https://link.aps.org/doi/10.1103/PhysRevB.106.024202} {\it  Phys. Rev. B} \textbf{106}, 024202 (2022).}

\bibitem{xu2021non}{Z. Xu, X. Xia, and S. Chen, Non-Hermitian Aubry-André model with power-law hopping, \color{blue} \href {https://link.aps.org/doi/10.1103/PhysRevB.104.224204} {\it  Phys. Rev. B} \textbf{104}, 224204 (2021).}

\bibitem{zhou2022topological}{L. Zhou and J. Gu, Topological delocalization transitions and mobility edges in the nonreciprocal Maryland model, \color{blue} \href {https://iopscience.iop.org/article/10.1088/1361-648X/ac4530/meta?casa_token=Tsv0mKGdBjIAAAAA:xMpLDRNPBr8fGwTsDbOOKPo8ULm-CqK8WUfIo4iwuqw4Mfv2FUbTdmEAhsXqyJPH4ZEeBgf1iuPvYkI-r6YXHvlr9kKE} {\it  J. Phys.: Condens. Matter} \textbf{34}, 115402 (2022).}

\bibitem{peng2023power}{D. Peng, S. Cheng, and G. Xianlong, Power law hopping
of single particles in one-dimensional non-Hermitian quasicrystals, \color{blue} \href {https://link.aps.org/doi/10.1103/PhysRevB.107.174205} {\it  Phys. Rev. B} \textbf{107}, 174205 (2023).}

\bibitem{fastenrath1990evidence}{U. Fastenrath, Evidence for Anderson transitions in 2D, \color{blue} \href {https://doi.org/10.1016/0038-1098(90)90642-O} {\it  Solid State Commun.} \textbf{76}, 855 (1990).}

\bibitem{white2020observation}{D. H. White, T. A. Haase, D. J. Brown, M. D. Hoogerland, M. S. Najafabadi, J. L. Helm, C. Gies, D. Schumayer, and D. A. Hutchinson, Observation of two-dimensional Anderson localisation of ultracold atoms, \color{blue} \href {https://doi.org/10.1038/s41467-020-18652-w} {\it  Nat. Commun.} \textbf{11}, 4942 (2020).}

\bibitem{bordia2017probing}{P. Bordia, H. Lüschen, S. Scherg, S. Gopalakrishnan, M. Knap, U. Schneider, and I. Bloch, Probing Slow Relaxation and Many-Body Localization in Two-Dimensional Quasiperiodic Systems, \color{blue} \href {https://link.aps.org/doi/10.1103/PhysRevX.7.041047} {\it  Phys. Rev. X} \textbf{7}, 041047 (2017).}

\bibitem{gautier2021strongly}{R. Gautier, H. Yao, and L. Sanchez-Palencia, Strongly Interacting Bosons in a Two-Dimensional Quasicrystal Lattice, \color{blue} \href {https://link.aps.org/doi/10.1103/PhysRevLett.126.110401} {\it  Phys. Rev. Lett.} \textbf{126}, 110401 (2021).}

\bibitem{szabo2020mixed}{A. Szabó and U. Schneider, Mixed spectra and partially extended states in a two-dimensional quasiperiodic model, \color{blue} \href {https://link.aps.org/doi/10.1103/PhysRevB.101.014205} {\it  Phys. Rev. B} \textbf{101}, 014205 (2020).}

\bibitem{vstrkalj2022coexistence}{A. Štrkalj, E. V. H. Doggen, and C. Castelnovo, Coexistence of localization and transport in many-body two-dimensional Aubry-André models, \color{blue} \href {https://link.aps.org/doi/10.1103/PhysRevB.106.184209} {\it  Phys. Rev. B} \textbf{106}, 184209 (2022).}

\bibitem{wang2023two}{Y. Wang, L. Zhang, Y. Wan, Y. He, and Y. Wang, Twodimensional vertex-decorated Lieb lattice with exact mobility edges and robust flat bands, \color{blue} \href {https://link.aps.org/doi/10.1103/PhysRevB.107.L140201} {\it  Phys. Rev. B} \textbf{107}, L140201 (2023).}

\bibitem{lieb1989two}{E. H. Lieb, Two Theorems on the Hubbard Model, \color{blue} \href {https://link.aps.org/doi/10.1103/PhysRevLett.62.1201} {\it  Phys. Rev. Lett.} \textbf{62}, 1201 (1989).}

\bibitem{mukherjee2015observation}{S. Mukherjee, A. Spracklen, D. Choudhury, N. Goldman, P. Öhberg, E. Andersson, and R. R. Thomson, Observation of a Localized Flat-Band State in a Photonic Lieb Lattice, \color{blue} \href {https://link.aps.org/doi/10.1103/PhysRevLett.114.245504} {\it  Phys. Rev. Lett.} \textbf{114}, 245504 (2015).}

\bibitem{vicencio2015observation}{R. A. Vicencio, C. Cantillano, L. Morales-Inostroza, B. Real, C. Mejía-Cortés, S. Weimann, A. Szameit, and M. I. Molina, Observation of Localized States in Lieb Photonic Lattices, \color{blue} \href {https://link.aps.org/doi/10.1103/PhysRevLett.114.245503} {\it  Phys. Rev. Lett.} \textbf{114}, 245503 (2015).}

\bibitem{diebel2016conical}{F. Diebel, D. Leykam, S. Kroesen, C. Denz, and A. S. Desyatnikov, Conical Diffraction and Composite Lieb Bosons in Photonic Lattices, \color{blue} \href {https://link.aps.org/doi/10.1103/PhysRevLett.116.183902} {\it Phys. Rev. Lett.} \textbf{116}, 183902 (2016).}

\bibitem{xia2018unconventional}{S. Xia, A. Ramachandran, S. Xia, D. Li, X. Liu, L. Tang, Y. Hu, D. Song, J. Xu, D. Leykam, S. Flach, and Z. Chen, Unconventional Flatband Line States in Photonic Lieb Lattices, \color{blue} \href {https://link.aps.org/doi/10.1103/PhysRevLett.121.263902} {\it  Phys. Rev. Lett.} \textbf{121}, 263902 (2018).}

\bibitem{taie2015coherent}{S. Taie, H. Ozawa, T. Ichinose, T. Nishio, S. Nakajima, and Y. Takahashi, Coherent driving and freezing of bosonic matter wave in an optical Lieb lattice, \color{blue} \href {https://www.science.org/doi/10.1126/sciadv.1500854} {\it  Sci. Adv.} \textbf{1}, e1500854 (2015).}

\bibitem{baboux2016bosonic}{F. Baboux, L. Ge, T. Jacqmin, M. Biondi, E. Galopin, A. Lemaître, L. Le Gratiet, I. Sagnes, S. Schmidt, H. E. Türeci, A. Amo, and J. Bloch, Bosonic Condensation and DisorderInduced Localization in a Flat Band, \color{blue} \href {https://link.aps.org/doi/10.1103/PhysRevLett.116.066402} {\it Phys. Rev. Lett.} \textbf{116}, 066402 (2016).}

\bibitem{slot2017experimental}{M. R. Slot, T. S. Gardenier, P. H. Jacobse, G. C. P. van Miert, S. N. Kempkes, S. J. M. Zevenhuizen, C. M. Smith, D. Vanmaekelbergh, and I. Swart, Experimental realization and characterization of an electronic Lieb lattice, \color{blue} \href {https://www.nature.com/articles/nphys4105} {\it Nat. Phys.} \textbf{13}, 672 (2017).}

\bibitem{goda2006inverse}{M. Goda, S. Nishino, and H. Matsuda, Inverse Anderson Transition Caused by Flatbands, \color{blue} \href {https://link.aps.org/doi/10.1103/PhysRevLett.96.126401} {\it Phys. Rev. Lett.} \textbf{96}, 126401 (2006).}

\bibitem{nishino2007flat}{S. Nishino, H. Matsuda, and M. Goda, Flat-band localization in weakly disordered system, \color{blue} \href {https://doi.org/10.1143/JPSJ.76.024709} {\it J. Phys. Soc. Jpn.} \textbf{76}, 024709 (2007).}

\bibitem{chalker2010anderson}{J. T. Chalker, T. S. Pickles, and P. Shukla, Anderson localization in tight-binding models with flat bands, \color{blue} \href {https://link.aps.org/doi/10.1103/PhysRevB.82.104209} {\it Phys. Rev. B} \textbf{82}, 104209 (2010).}

\bibitem{bodyfelt2014flatbands}{J. D. Bodyfelt, D. Leykam, C. Danieli, X. Yu, and S. Flach, Flatbands under Correlated Perturbations, \color{blue} \href {https://link.aps.org/doi/10.1103/PhysRevLett.113.236403} {\it Phys. Rev. Lett.} \textbf{113}, 236403 (2014).}

\bibitem{leykam2017localization}{D. Leykam, J. D. Bodyfelt, A. D. Desyatnikov, and S. Flach, Localization of weakly disordered flat band states, \color{blue} \href {https://doi.org/10.1140/epjb/e2016-70551-2} {\it Eur. Phys. J. B} \textbf{90}, 1 (2017).}

\bibitem{roy2020interplay}{N. Roy, A. Ramachandran, and A. Sharma, Interplay of disorder and interactions in a flat-band supporting diamond chain, \color{blue} \href {https://link.aps.org/doi/10.1103/PhysRevResearch.2.043395} {\it Phys. Rev. Res.} \textbf{2}, 043395 (2020).}

\bibitem{ahmed2022flat}{A. Ahmed, A. Ramachandran, I. M. Khaymovich, and A. Sharma, Flat band based multifractality in the all-band-flat diamond chain, \color{blue} \href {https://link.aps.org/doi/10.1103/PhysRevB.106.205119} {\it Phys. Rev. B} \textbf{106}, 205119 (2022).}

\bibitem{lee2023critical}{S. Lee, A. Andreanov, and S. Flach, Critical-to-insulator transitions and fractality edges in perturbed flat bands, \color{blue} \href {https://link.aps.org/doi/10.1103/PhysRevB.107.014204} {\it Phys. Rev. B} \textbf{107}, 014204 (2023).}

\bibitem{shklovskii1993statistics}{B. I. Shklovskii, B. Shapiro, B. R. Sears, P. Lambrianides, and H. B. Shore, Statistics of spectra of disordered systems near the metal-insulator transition, \color{blue} \href {https://link.aps.org/doi/10.1103/PhysRevB.47.11487} {\it Phys. Rev. B} \textbf{47}, 11487 (1993).}

\bibitem{oganesyan2007localization}{V. Oganesyan and D. A. Huse, Localization of interacting fermions at high temperature, \color{blue} \href {https://link.aps.org/doi/10.1103/PhysRevB.75.155111} {\it Phys. Rev. B} \textbf{75}, 155111 (2007).}

\bibitem{lindblad1976generators}{G. Lindblad, On the generators of quantum dynamical semigroups, \color{blue} \href {https://doi.org/10.1007/BF01608499} {\it Commun. Math. Phys.} \textbf{119}, 48 (1976).}

\bibitem{gorini1976completely}{V. Gorini, A. Kossakowski, and E. C. Sudarsahan, Completely positive dynamical semigroups of $N$-evel systems, \color{blue} \href {https://doi.org/10.1007/BF01608499} {\it J. Math. Phys.} \textbf{17}, 821 (1976).}

\bibitem{dalibard1992wave}{J. Dalibard, Y. Castin, and K. Mølmer, Wave-Function Approach to Dissipative Processes in Quantum Optics, \color{blue} \href {https://link.aps.org/doi/10.1103/PhysRevLett.68.580} {\it Phys. Rev. Lett.} \textbf{68}, 580 (1992).}

\bibitem{carmichael1993quantum}{H. J. Carmichael, Quantum Trajectory Theory for Cascaded Open Systems, \color{blue} \href {https://link.aps.org/doi/10.1103/PhysRevLett.70.2273} {\it Phys. Rev. Lett.} \textbf{70}, 2273 (1993).}

\bibitem{ashida2020non}{Y. Ashida, Z. Gong, and M. Ueda, Non-Hermitian physics, \color{blue} \href {https://doi.org/10.1080/00018732.2021.1876991} {\it Adv. Phys.} \textbf{69}, 3 (2020).}

\bibitem{li2022engineering}{T. Li, Y.-S. Zhang, and W. Yi, Engineering dissipative quasicrystals, \color{blue} \href {https://link.aps.org/doi/10.1103/PhysRevB.105.125111} {\it Phys. Rev. B} \textbf{105}, 125111 (2022).}

\end{thebibliography}
\end{document}